\documentclass[journal]{vgtc}                     

\onlineid{0}

\vgtccategory{Research}

\vgtcpapertype{essay}

\title{A Priest, a Rabbi, and an Atheist Walk Into an Error Bar: \\ Religious Meditations on Uncertainty Visualization}

\author{%
  \authororcid{Michael Correll}{0000-0001-7902-3907} and
  \authororcid{Lane Harrison}{0000-0003-3029-2799}
}

\authorfooter{
  \item
  	Michael Correll is with Northeastern University's Roux Institute.
  	E-mail: m.correll@northeastern.edu.
  \item
  	Lane Harrison is with Worcester Polytechnic Institute.
  	E-mail: ltharrison@wpi.edu.
}

\abstract{
In this provocation, we suggest that much (although not all) current uncertainty visualization simplifies the myriad forms of uncertainty into error bars around an estimate. This apparent simplification into error bars comes only as a result of a vast metaphysics around uncertainty and probability underlying modern statistics. We use examples from religion to present alternative views of uncertainty (metaphysical or otherwise) with the goal of enriching our conception of what kind of uncertainties we ought to visualize, and what kinds of people we might be visualizing those uncertainties for.
}


\teaser{
  \centering
  \includegraphics[width=0.65\linewidth, alt={An angelic set of error bars}]{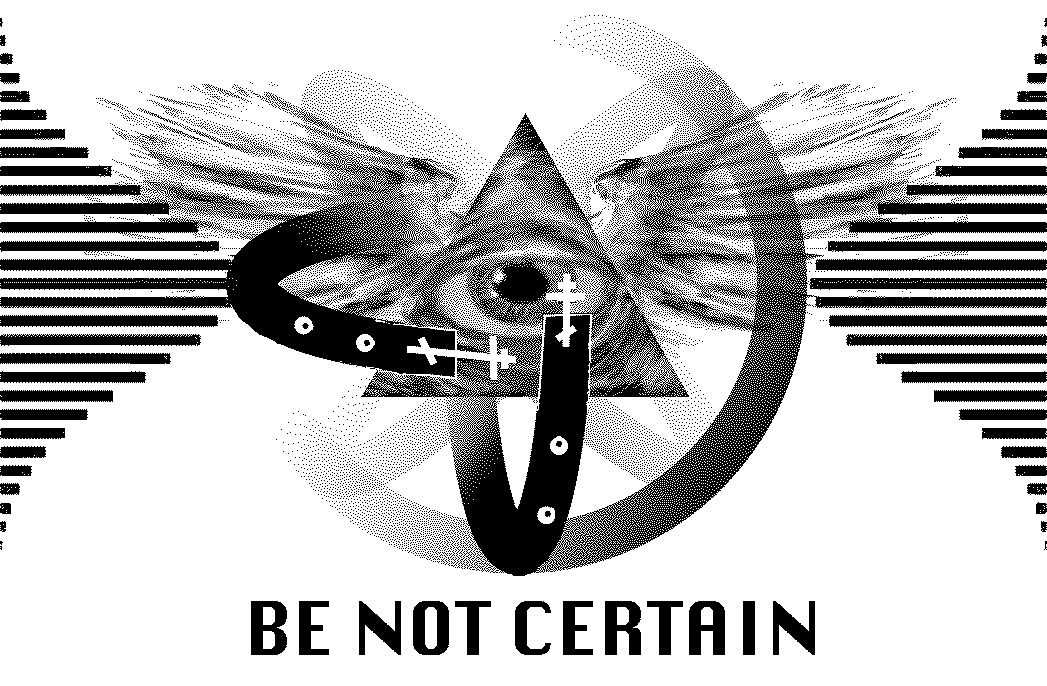}
  \label{fig:teaser}
}

\graphicspath{{figs/}{figures/}{pictures/}{images/}{./}} 

\usepackage{mathptmx}                  

\newcommand{\saw}{(\includegraphics[width=1em]{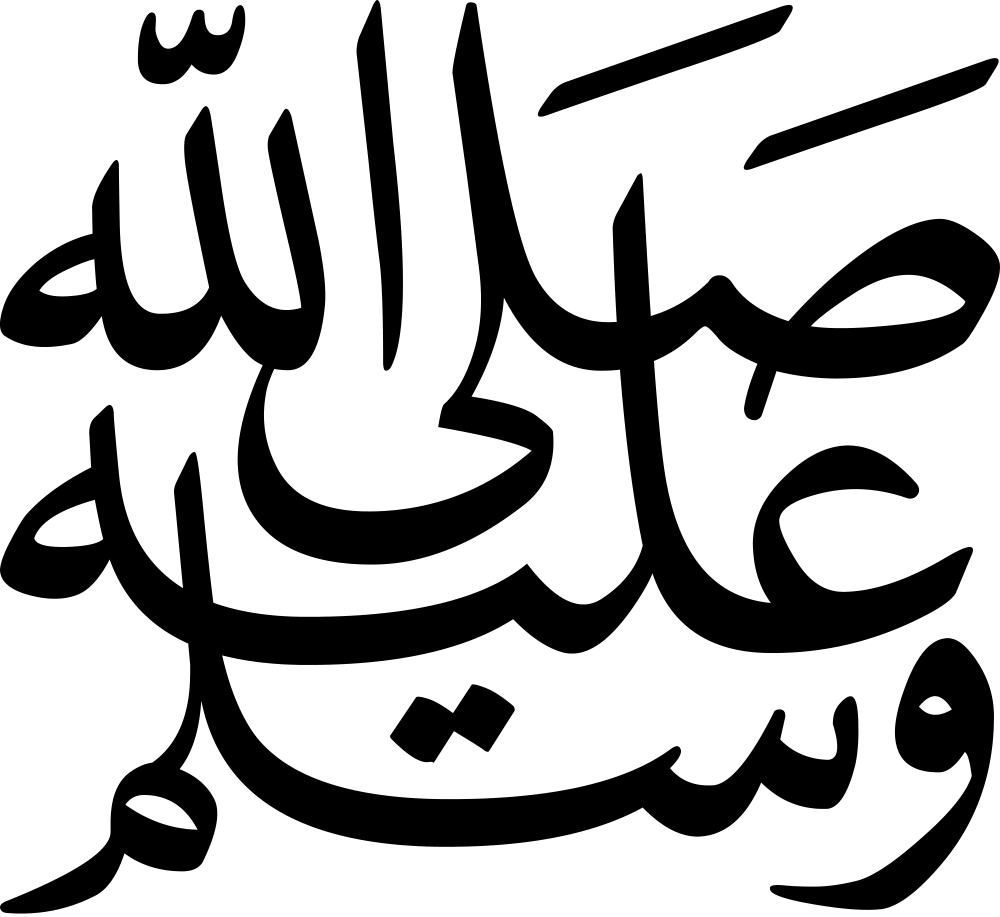})}

\begin{document}


\firstsection{Introduction}

\maketitle

\begin{quote}
\textit{We are talking about God; so why be surprised if you cannot grasp it? I mean, if you can grasp it, it isn't God. Let us rather make a devout confession of ignorance, instead of a brash profession of knowledge.}
\end{quote}
\hspace{3em}---St. Augustine, Sermon 117~\cite{hill1990works}\\

Uncertainty communication is one of a few examples (along with tax codes and most software) where an incredibly complicated assemblage of kludged-together mathematical concepts is presented to mass audiences, who are then blamed and belittled for not understanding them correctly. Take, for example, the error bar. The error bar looks superficially simple: some point estimate represented as a dot or a bar, and then some lines or brackets indicating some uncertainty in that estimate. Yet, the actual interpretation and use of the resulting interval is a matter of ongoing, quasi-philosophical debate among statisticians~\cite{joque2022revolutionary} (Bayesian credible intervals and frequentist confidence intervals have dramatically different meanings even when they produce the same numerical interval). What an error bar actually \textit{means} is esoteric enough that misinterpretations and misstatements are ubiquitous not just among statisticians and their students~\cite{belia2005researchers,hoekstra2014robust} but also text books~\cite{brewer1985behavioral}. Important information, necessary for using error bars for inferential purposes~\cite{cumming2005inference} (such as sample size, Bayesian priors, bootstrapping procedures, or even just a text description of what the error bars actually encode), is habitually absent~\cite{Correll2014} in scientific papers.
Visual judgments about the size, certainty, and reliability of effects is often variable, biased~\cite{Correll2014,newburger2022fitting,sarma2024odds}, and based on ``satisficing''~\cite{kale2020visual} strategies that elide relevant information (like sample size\cite{fouriezos2008visual}) and so may have very little to do with the inferential statistics at play. The statistical forms of uncertainty may themselves have little to do with the richer, human-centered forms of uncertainty: the often qualitative ``implicit errors''~\cite{panagiotidou2021implicit} and ``data hunches''~\cite{lin2022data} about data that spill beyond the boundaries of a single error bar.

Given all of these deficiencies, why, then, are error bars ubiquitous in scientific work? Because they are \textit{rituals}~\cite{gigerenzer2018statistical}: quasi-superstitious practices based on---often unexamined---philosophical underpinnings (c.f., Brown and Swift~\cite{browne2018other} on how the algorithms of artificial intelligence often have similar ritual trappings). We include error bars because they are expected of us, to appease an unknown body of reviewers, or to add appropriate solemnity or ``science-ness'' to the data we collect. The uncertainty is calculated even if the utility of presenting the uncertainty information is dubious~\cite{hullman2019authors}, or the statistical interpretation nonsensical or valueless. Take for example an anecdote from Cohen's~\cite{cohen1994earth} critique of null-hypothesis statistics testing:

\begin{quote}
\textit{A colleague approaches me with a statistical problem. He believes that a generally rare disease does not exist at all in a given population, hence $H_0$: $P = 0$. He draws a more or less random sample of 30 cases from this population and finds that one of the cases has the disease, hence $P_s$ = $1/30$ = $.033$. He is not sure how to test $H_0$, chi-square with Yates's correction or the Fisher exact test, and wonders whether he has enough power. Would you believe it? And would you believe that if he tried to publish this result without a significance test, one or more reviewers might complain?}
\end{quote}

Of course, the ``right'' answer is that \textit{no} statistical test is needed: the colleague assumed that a disease was not present in a population, but discovered that, in fact, it \textit{was}, and so the original hypothesis was definitively falsified: there is no inferential uncertainty to be quantified or presented, nor would an error bar be relevant to present this information. Yet, the urge to perform the ritual remains. This is not to say that rituals are never useful: many rituals come with many benefits, even if only psychological, and statistical rituals are no exception. But the particular statistical philosophies underpinning rituals like error bars are often relative newcomers (for instance, the t-distribution underlying frequentist t-confidence interval is almost exactly 150 years old~\cite{pfanzagl1996studies}, and the computational advances that make Bayesian credible intervals feasible younger still~\cite{joque2022revolutionary}), and yet there are often quasi-moralist judgments associated with the ``failure'' or lack of comprehension of these rituals, and the assumption that there exists a sainted elect who are the \textit{real} intercessors towards statistical truth, forever apart from the flawed masses subject to insurmountable cognitive biases.

Core to the operation of modern statistics, and reified in the ``correct'' interpretation and use of uncertainty visualizations, are philosophical underpinnings that often have more of the trappings of faith than of science. Before we critique the viewers of our uncertainty visualizations for misinterpreting our designs, or making ``biased'' or seemingly ``contradictory'' decisions not in keeping with the ``homo economicus''~\cite{urbina2019critical} of ``rational'' decision theory, we should examine our own parochial and blinkered views. We echo Berkeley~\cite{berkeley1754analyst} who, in his famous critique of the notion of infinitesimals in calculus, asks:
\begin{quote}
\textit{Whether Mathematicians, who are so delicate in religious Points, are strictly scrupulous in their own Science? Whether they do not submit to Authority, take things upon Trust, believe Points inconceivable? Whether they have not their Mysteries, and what is more, their Repugnancies and Contradictions?} 
\end{quote}

In this paper, we look to an older, and more human-centered, set of rituals around uncertainty: that of religious attitudes towards doubt, uncertainty, and confidence. Dealing with uncertainty and doubt is perhaps one of the key functions of religion: per Hebrews 11:1, ``Now faith is the substance of things hoped for, the evidence of things not seen.'' Religious views towards faith, evidence, jurisprudence, and certainty have had centuries of explication and development, and are inescapably wrapped up in the genesis of our current mathematical conceptions of probability and evidence~\cite{franklin2015science}. We present a series of vignettes about religious attitudes towards uncertainty that conflict, interrogate, or in some cases supersede the statistical views that underlie uncertainty visualizations like error bars. We do so as a provocation for our community to rethink our unquestioned assumptions around uncertainty as well as the limitations in our techniques for communicating the unknown.

\subsection{Positionality and Limitations}
For reasons of both positionality and experience, we focus on Abrahamic religions in this work. That being said, neither of the authors are theologians. We can only ask forgiveness for any omissions or misstatements or omissions. Per Psalms 19:14 ``Let the words of my mouth, and the meditation of my heart, be acceptable in thy sight, O Lord, my strength, and my redeemer.''

Likewise, entire works have focused on just single aspects of some of the religious and mathematical concepts we discuss in this paper. Space restrictions and our own lack of expertise limit us to only a superficial gloss over these matters. This work is meant to be primarily exhortatory and an instigator of reflection rather than a full and complete explication of the matters at stake.
As per a hadith attested by Ammar bin Yasir, the Prophet \saw{} said: ``Make your prayer long and your sermon short'' (Riyad as-Salihin 699).

\section{On the Unknowable}
Error bars are used (often erroneously) as tools for visually assessing uncertainty around an estimate or around the robustness and reliability of the difference between two samples means~\cite{cumming2005inference}. Yet, these visual uncertainties are based on an information that is unknowable: say, the exact value of a population mean, or the unreachable asymptotic convergence of some process of sampling or observation. These uncertainties are also often uninteresting: for instance, the commonality of what Cohen~\cite{cohen1994earth} calls the ``nil'' hypothesis: two populations are certain to have population means that differ at least \textit{a little} (even if it's quite a few decimal places down), and so the common ``null hypothesis'' that two populations means are identical is almost always false. If a core aspect of Popperian science is falsifiability~\cite{popper1963science}, then an error bar is a poor tool: it rarely falsifies anything, or, at least, anything useful.

The issue of how to learn useful things about matters that are fundamentally unknowable is not unique to statistics: The nature of God is also often conceived of as unknowable, relying as it does on finite and limited human perspective to grasp the infinite and unlimited nature of the Divine (Ecclesiastes 5:2 ``Be not rash with thy mouth, and let not thine heart be hasty to utter any thing before God: for God is in heaven, and thou upon earth: therefore let thy words be few.''). Maimonides~\cite{maimonides2012guide} suggests that ordinary procedures of seeking positive knowledge is mistaken when considering God, but that instead it is more fruitful to establish proof by \textit{negation}:

\begin{quote}
\textit{It will now be clear to you, that every time you establish by proof the negation of a thing in reference to God, you become more perfect, while with every additional positive assertion you follow your imagination and recede from the true knowledge of God.}
\end{quote}

That is, to attempt to use ordinary logic to determine if God has weight or mass or the ability to see is something of a category error. But, by determining what God is \textit{not}, we can eventually arrive closer to an understanding of what God \textit{is}. Even just listing positive properties of God may be a form of insult, since such lists are necessarily incomplete. Here Maimonides cites a section from the Talmud (Berakhot 33b):

\begin{quote}
\textit{It is comparable to a king who possessed many thousands of golden dinars, yet they were praising him for silver ones. Isn’t that deprecatory? All of the praises we could possibly lavish upon the Lord are nothing but a few silver dinars relative to many thousands of gold dinars. Reciting a litany of praise does not enhance God's honor.}
\end{quote}

Future theological works take this concept of knowledge by negation even further. For instance, the 14th century anonymous work of Christian mysticism \textit{The Cloud of Unknowing}~\cite{wolters2018cloud} holds that positive knowledge of God through the intellect is impossible and prone to error (or hubris), and so proposes contemplation through a ``Cloud of Forgetting'': an intentional clearing of the mind and disconnection from worldly affairs that brings one into the titular ``cloud of unknowing'', the place where, perhaps somewhat paradoxically, true divine wisdom is possible. This notion of coming to know God through negation is a fundamentally different procedure for knowledge than positive forms of, say, measuring or sampling objects in the world.  

\begin{figure}
    \centering
    \includegraphics[width=0.95\columnwidth]{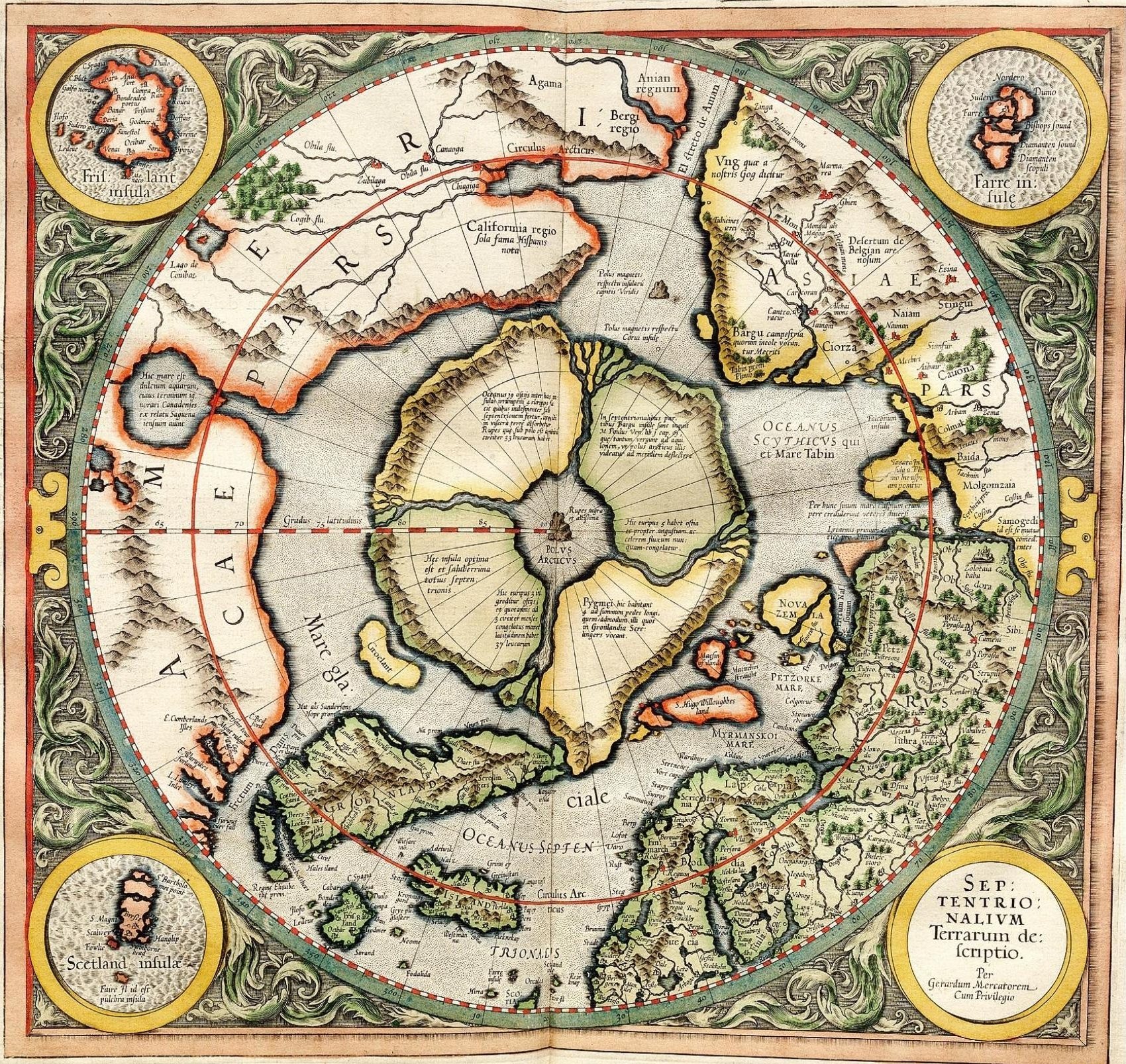}
    \caption{Posthumously published map by Gerhard Mercator of the Arctic, 1595. The then-unexplored North Pole is neither blank nor filled with ``here there be dragons'' marks of uncertainty: rather, assumptions derived from Biblically-powered ``sacred geography''~\protect\cite{gravon2024gerhard} fill the space. Public domain, image from \href{https://commons.wikimedia.org/wiki/File:Mercator_north_pole_1595.jpg}{Wikimedia commons}.
    }
    \label{fig:mercator}
\end{figure}

As a final anecdote on religious attitudes towards the unknown, we focus on an example of how religious knowledge can function as a way to fill in gaps in empiricism, or produce strong priors that can inform how unknown data are visualized. Gravon~\cite{gravon2024gerhard} uses the example of a map that Mercator made of the north pole (\autoref{fig:mercator}). At the time, polar expeditions were few and poorly documented, and the unreliability of instrumentation created discrepancies between the magnetic and geographic north poles. Per Gravon, ``Not only was Mercator’s empirical data on
magnetic north precarious, but his supporting historical source on the region’s
topography was also dubious.'' In the absence of knowledge, and facing a desire to ``project certitude'', Mercator turned to a ``\textit{prima scriptura}'' approach. Rather than a lacuna with periodic sea monsters (as in other maps of unknown regions and, indeed, in future versions of polar regions copied from Mercator's version), Mercator filled his map with geography inspired by a contemporary theory equating the north pole with biblical paradise. This anecdote suggest to us both that uncertainty is rarely if ever \textit{total} (there are often priors, and very specific priors, that the analyst has in mind), but that also that making these priors explicit may have benefits, as in Kim et al.~\cite{kim2017explaining} work where eliciting predictions (and so affording the ability to see how the actual data did or did not surprisingly differ from such predictions) improved users' understanding and  recall of data.

We present these anecdotes to suggest two important philosophical considerations in uncertainty visualization. The first is whether our techniques are overly focused on attempting to capture \textit{positive} knowledge (for instance, an estimate of a value with some associated precision) rather than the often just as important \textit{negative} knowledge (What remains unknown? What properties are we confident that our data does \textit{not} have?). The second is the importance of of prior knowledge, both in the form of the typical absence of literal Bayesian-style priors in uncertainty visualizations, but also of the absence of higher level information like the many critical ``entanglements''~\cite{akbaba2024entanglements} in our work, or even just the epistemologies underlying what we consider to be data or evidence.



\section{On the Casting of Lots}
One of the key components that led to the modern Bayesian revolution in statistics was the availability of computational power to conduct to make practical the calculations needed for Bayesian methods~\cite{joque2022revolutionary}. These calculations are mostly spent on simulations, shuffles, and other forms of ``Monte Carlo'' style randomness (such as bootstrapping~\cite{efron1992bootstrap}, or MCMC~\cite{robert2011short}). When we use these random methods to make judgments and predictions, we are in effect performing a computationally more complicated process of \textit{casting lots}. Religions have had contentious relationships with these forms of chance-based divination.

For instance, the Qu'ran explicitly forbids the casting of lots for divinatory purposes. Instead, in several hadiths, an alternative prayer (the Salat al-Istikharah) is suggested (e.g., Mishkat al-Masabih 1323):

\begin{quote}
\textit{God's Messenger \saw{} used to teach us how to ask God’s guidance about matters just as he used to teach as a sura of the Qur'an, saying: When any of you intends to do something he should pray two rak'as which are not compulsory, then say, ``O God, I ask Thy guidance by Thy knowledge, I ask Thee for power by Thy power, and I ask Thee out of Thy great abundance, for Thou hast power and I have none, Thou knowest and I did not, and Thou art the One who is aware of the unseen. O God, if Thou knowest that this matter is good for me regarding my religion, my livelihood, and my future wellbeing (or he said, ``my affairs in this world and the next''), ordain it for me and make it easy for me, then bless me in it. But if Thou knowest that this matter is bad for me regarding my religion, my livelihood, and my future wellbeing (or he said, ``my affairs in this world and the next''), turn it away from me, turn me away from it, ordain good for me wherever it is, then make me pleased with it.” He said that he should name what he required.}
\end{quote}

While both casting lots and the Istikharah can involve external forces (for instance, a thing being ``made easy'' functioning as a sign that it is the correct path), the focus on internal reflection and exercising moral judgment distinguishes it from pure chance. The goal is to clarify, not choose.


The Bible also forbids the casting of lots for divination (Deuteronomy 18:10). However, the casting of lots for discernment is permitted. Divination can be thought of as seeking hidden knowledge about the future, whereas discernment involves decisions where logic, consensus, or other forms of guidance prove insufficient. In the story of Jonah, the titular prophet is asked by God to travel to the city of Nineveh, to publicly call out the wickedness of its inhabitants. Jonah instead flees by ship, which is threatened by a sudden storm. Lots are cast by the sailors to ascertain the cause of their misfortune:

\begin{quote}
\textit{And they said every one to his fellow, Come, and let us cast lots, that we may know for whose cause this evil is upon us. So they cast lots, and the lot fell upon Jonah.\\--- Jonah 1:7
\\\\
And he said unto them, Take me up, and cast me forth into the sea; so shall the sea be calm unto you: for I know that for my sake this great tempest is upon you.\\--- Jonah 1:12}
\end{quote}

Despite many instances of the casting of lots in the Bible-- ``The lot is cast into the lap; but the whole disposing thereof is of the Lord'' (Proverbs 16:33)-- specific artifacts and methods for casting lots remain unknown. One example are  the divinatory Urim and Thummin. Part of the high priest's vestments as laid out in Exodus 28:30, the Urim and Thummim are a set of small marked objects drawn from a pouch. While the artifact is named, the method of use is curiously absent. Speculation about methodologies range from the divine: with stones glowing to indicate a response, to mechanistic: systematically drawing stones to formulate a response. 

Randomness and sheer chance remains a core component of uncertainty communication. For instance, with all the alleged authority of error bars, sampling error alone is sufficient to have them ``dance''~\cite{dragicevic2016fair} wildly across criteria of significance or effect size. Our examples suggest both practical and moral hazards when one places a heavy reliance on chance in the absence of moral judgment.

\section{On Evidence}
Error bars implicitly or explicitly rely on a notion of confidence. For instance, the $\alpha=0.95$ significance level, connected with the all-important $p<0.05$ measure of significance, is a common (and likely modal) parameter setting in both Bayesian credible intervals and frequentist confidence intervals. While the history and justification of the $p=0.05$ threshold is long and winding~\cite{cowles1982origins}, this history can be summarized by Fisher saying ``Personally, the writer prefers to set a low standard of significance at the 5 per cent point, and ignore entirely all results which fail to reach this level'' in 1926, and the rest of the field following suit. Religious standards of certainty are much more varied and have much longer historical footprints.


For instance, after a discussion of the high stakes and importance of criminal cases where the death penalty is on the table, the Talmud (Sanhedrin.37b.2) imposes a very high standard upon the certainty of witnesses:
\begin{quote}
    \textit{How does the court describe testimony based on conjecture? The court says to the witnesses: Perhaps you saw this man about whom you are testifying pursuing another into a ruin, and you pursued him and found a sword in his hand, dripping with blood, and the one who was ultimately killed was convulsing. If you saw only this, it is as if you saw nothing, and you cannot testify to the murder.}
\end{quote}

Even divine revelation is not always sufficient evidence. For instance, in an anecdote recounted in the Talmud (Bava Metzia 59b), Rabbi Eliezer produces various miracles in support of his interpretation of \textit{halakhic} law, culminating in a voice from the heavens saying ``Why are you differing with Rabbi Eliezer, as the \textit{halakha} is in accordance with his opinion in every place that he expresses an opinion?''. But the other rabbis do not view this miracle as sufficient proof, as this does not constitute a \textit{majority} opinion; upon hearing this, God then smiles and responds ``My children have triumphed over Me; My children have triumphed over Me.''

Another Talmudic concept of uncertainty (and often counter-intuitive, especially for those with backgrounds in modern conceptions of probability and decision-theory), is the distinction between cases involving uncertainty information that are deemed ``\textit{parish}'' or ``\textit{kavu'a}''~\cite{kazhdan2018logical}. A common example of this distinction is a town which has ten butchers, nine of which are kosher (Pesachim 9b):
\begin{quote}
\textit{With regard to nine stores in a city, all of which sell kosher meat from a slaughtered animal, and one other store that sells meat from unslaughtered animal carcasses, and a person took meat from one of them and he does not know from which one he took the meat, in this case of uncertainty, the meat is prohibited. This ruling is based on the principle: The legal status of an item fixed in its place is that of an uncertainty that is equally balanced. In this case, when it comes to determining whether or not this meat comes from a kosher store, the two types of stores are regarded as though they were equal in number.}
[...]
\textit{And in the case of meat found outside, follow the majority. If most stores in the city sell kosher meat one can assume that the meat he found is kosher, based on the principle: Any item separated, i.e., not fixed in its place, is presumed to have been separated from the majority.}
\end{quote}

That is, while in both cases the probability of the meat being kosher would seem to be the same (90\%), in the case where one has forgotten which butcher one purchased it from is seen as having two equally likely outcomes (\textit{kavu'a}): either purchased from a kosher butcher or not, and this 50\% standard is insufficient to deem the meat as kosher. Whereas, if the same meat is encountered \textit{in situ}, rather than brought away, then the ``majority rules'' (\textit{parish}), and the meat is deemed kosher (importantly, not just ``probably kosher'', but \textit{definitively} so). Determining which decision problems fall into which category, and then aligning those decisions to contemporary understandings of probability (as in Koppel~\cite{koppel2003resolving}), is complicated and often seemingly counter-intuitive, and involves grappling with many Talmudic edge cases. But the central insight is that context can (re-)define hypothesis spaces and the very meaning of probabilities.

An example of the impact of this failure to consider reference classes and hypothesis spaces occurs in visualizing uncertainty in weather data. For instance, icon arrays have seen sporadic use in contexts like communicating medical risk, but there the meaning of an individual icon can be relatively straightforward: one icon could represent a sample patient, who is either sick or healthy, for instance (as in Ottley et al.~\cite{ottley2015improving}). However, for icon arrays of weather data, is an icon a single run of the model, or a portion of the forecast area, or a representation of a day similar to the one being forecasted? Trying to understand the reference class behind weather forecasts remains a key challenge in uncertainty communication~\cite{stephens2012communicating}. Regardless of their mathematical equivalences, not all probabilities are interpreted in the same way by the same people.







\section{Discussion \& Conclusion}


Statistical probabilities are not properties of the natural world--- there are not wild untamed packs of p-values roaming the plains. Rather, statistics and probability are ways of interpreting the world, subject to a process of real abstraction~\cite{joque2022revolutionary} that, for instance, can strip context from observations, lend unearned objectivity or perceived neutrality to very particular agendas, or used to justify collecting more and more data from increasingly atomized and disconnected individuals so they can be categorized and modeled and predicted. While many scientific fields draw on and prosper from these foundations, it is important to contextualize them as relatively recent in the whole of human history, and as part of a value system that is not universally held or experienced in the same way. 
In this paper we contrast some of the modern abstractions surrounding randomness and uncertainty with older ones, ones which still deeply shape how people experience the world and engage with each other.

Doubt and uncertainty are universal. Yet, we feel that many standards in uncertainty visualizations focus on a narrow set of rituals that capture only a small fraction of this experience. We beat our observations into specific statistical shapes, and then blame (or patronize) our users when these rituals fail to satisfy. Rituals serve sociological needs as well as psychological ones. They simplify by collapsing the complexities of uncertainty into manageable symbols and actions: an icon array, a pouch of stones, a prayer. They punish by denoting what is correct or acceptable, and shame or exclude those who transgress, even when justified. Are our statistical rituals then really so different? Anthropologist Mary Douglas~\cite{douglas2013risk} writes that ``Taboo works because a community of believers has developed a consensus ... They deal with their risks by mustering solidarity, invoking danger to maintain difference''.
A community following certain rituals, using specific statistical tests or error bars for example, might similarly invoke danger to warn against any perceived misuse, and levy consequences for missteps in order to maintain social cohesion.

The existence of diversity of thought does not permit the designer of uncertainty of visualizations a retreat to a na\"ive sort of nihilistic epistemic relativism. Just because there are many rituals associated with uncertainty, you are not rid of the burden of dutifully and rigorously performing the rituals that you have chosen, whether it is frequentist or Bayesian statistics, or more esoteric ways of grappling with the unknown. What we call for instead is an awareness and intentionality around these epistemic choices, and a realization that visualization designers are all too often immersed in a particular (and particularly narrow) view of uncertainty that has more to do with ritual and habit than rigorous and unquestionable statistical truth, and that assuming that such rituals are the default way of dealing with the unknown can leave us blinkered or parochial. Per Douglas~\cite{douglas2013risk}: ``The innocent view of culture is that we don’t have it at home; it is only abroad that people are culturally hide-bound. A special effort of sophistication is necessary to see our own culture.''




Most people in the world hold religious views. In our extremely cursory overview of just a fraction of these views, we have shown that there exist many different assumptions about uncertainty, evidence, and the forces that act upon them.  
A formal research agenda is beyond the scope of this work (and these authors), but we might take inspiration from traditions that emphasize respectful engagement with people of different backgrounds: ``Be not forgetful to entertain strangers: for thereby some have entertained angels unawares.'' (Hebrews 13:2). Designing visualizations for a diverse global community might require us acknowledge diverse ways of reasoning with uncertainty, and expand our language, design, and even metaphysics. If we do not, we risk a two-faced view of uncertainty visualization: one that assumes that we are designing either for a group that have already signed on for and fully understand the statistical underpinnings behind the uncertainty information we want to share, or a na\"ive and helpless audience of people who must be patronized to lest they misinterpret the uncertainty information we have deigned to show to them. In such a world, it is the visualization community that will be left behind: ``For we walk by faith, not by sight'' (II Corinthians 5:7).


\acknowledgments{%
We thank Jessica Hullman for comments on a draft of this work. This work was supported in part by a grant from NSF, \#2418908.%
}

\bibliographystyle{abbrv-doi-hyperref}

\bibliography{template}

\end{document}